\documentclass[a4paper]{spie}  
\usepackage[]{graphicx}
\usepackage{hyperref}
\include{aas_macros.tex}
\title{End to end numerical simulations of the MAORY multiconjugate adaptive optics system} 

\author{Arcidiacono C.\supit{a}, Schreiber L.\supit{a}, Bregoli G.\supit{a}, Diolaiti E.\supit{a}, Foppiani I.\supit{a}, Cosentino G.\supit{b}, Lombini M.\supit{a}, Butler R.~C.\supit{c} and Ciliegi P.\supit{a}
\skiplinehalf
\supit{a}INAF - Osservatorio Astronomico di Bologna, Via Ranzani 1, I-40127 Bologna, Italy; \\
\supit{b}Universit\'a degli Studi di Bologna, Dipartimento di Fisica e Astronomia, Viale Berti Pichat, 6/2 I-40127 Bologna, Italy;\\
\supit{c}INAF - Istituto di Astrofisica Spaziale e di Fisica Cosmica, Via Piero Gobetti 101, I-40129 Bologna, Italy
}

\authorinfo{Further author information: (Send correspondence to Carmelo Arcidiacono)\\Carmelo Arcidiacono: E-mail: carmelo.arcidiacono@oabo.inaf.it, Telephone: +39 051 209 5704}

\begin{document} 
  \maketitle 

\begin{abstract}
MAORY is the adaptive optics module of the E-ELT that will feed the MICADO
imaging camera through a gravity invariant exit port. MAORY has been
foreseen to implement MCAO correction through three high order deformable
mirrors driven by the reference signals of six Laser Guide Stars (LGSs)
feeding as many Shack-Hartmann Wavefront Sensors. A three Natural Guide
Stars (NGSs) system will provide the low order correction.
We develop a code for the end-to-end simulation of the MAORY adaptive optics (AO) system
in order to obtain high-ﬁdelity modeling of the system performance.
It is based on the IDL language and makes extensively
uses of the GPUs. Here
we present the architecture of the simulation tool and its achieved and
expected performance.\end{abstract}


\keywords{Multi-Conjugate Adaptive Optics, E-ELT, Laser Guide Stars, Sodium Layer, MAORY, Simulations}

\section{INTRODUCTION}
\label{sec:intro}  

The Multi conjugate Adaptive Optics RelaY (MAORY\cite{maory}) is the adaptive optics module of the
European Extremely Large Telescope (E-ELT\cite{E-ELT}) that will feed the Multi-AO
Imaging Camera for Deep Observations (MICADO) through a gravity invariant
exit port and a second possible instrument. MAORY has been foreseen to
implement Multi-Conjugate Adaptive Optics (MCAO) correction through three high order deformable mirrors
driven by the reference signal of six Laser Guide Stars (LGSs) feeding as
many Shack-Hartmann Wavefront Sensors. A three Natural Guide Stars (NGSs)
system will provide the low order correction.

We are developing a numerical simulation tool to accomplish the end to end (E2E)
MonteCarlo (MC) simulation. This code predicts the performance and allows to
fine tune the system characteristics of MAORY. It is based on the Interactive Data
Language (IDL) and it makes an extensively usage of graphics processing units (GPU).
Through the calls to C/C++ and CUDA routines we improve the speed
at the cost of a larger complexity for the memory management. See Section~\ref{sec:language}.

We designed
the code in order to be modular and adaptable: the different simulation steps, namely
the atmosphere generation, the open loop wave-front measurements,
the interaction matrix calibration, the closed loop, the Point Spread
Functions (PSFs) generation are performed independently. The results of
each of these steps is saved in files on the disk and these are the input
for the following module. See Section~\ref{sec:scheme}.

The code implements different wavefront Sensors (WFS) and both Natural and Laser Guide Stars (NGS and LGS) see Section~\ref{sec:components}. 

Finally some example for 8m class telescope is given Section~\ref{sec:examples}.

\section{Different Languages} 
\label{sec:language}
The previous experiences in numerical computing of the team members was the driver for the selection of the IDL language 
as base for the architecture of the code. We provided it of high level routines which take into account configurations, skeleton and loop. The low level ones provide the repetitive and mathematical jobs. Some of the low level routines of the STARFINDER\cite{2000Msngr.100...23D} and of the LOST\cite{2004ApOpt..43.4288A} numerical tools have been used and adapted to serve the new tool.

The E2E simulations of Extremely large telescopes (ELTs) require extreme performances to obtain results in an human acceptable time. The use of multi--core workstation is mandatory and the requirements on memory are tough: at least 128Gb memory RAM are needed for a full MCAO ELT simulation.
An E2E E-ELT configuration requires in the real action phase, when loop is simulated, to store in the RAM mirror modes, open loop wavefront (WF) arrays of the simulated Adaptive Optics (AO) references and test stars for Point Spread Function (PSF) and Strehl Ratio (SR) computation, control matrix, slope vectors and a few more service array and structure. For MAORY we typically consider a telescope pupil inscribed on 800$\times$800 elements for a 4.875e-2 meter per pixel, this value 
sets all the other important ones: considering $\approx 5000$  modes for the ground layer Deformable Mirror (DM), $\approx 1400$ and $\approx 1800$ respectively for the two post focal DMs with 1m 
 actuator pitch projected on the primary, a $3\times 3$ constellation of test stars, for 2seconds of run, the simulation needs about 100GB of available RAM.

Many routines of IDL are multi-threaded but some critical one are not, an example for all is the Singular Value Decomposition in its various form from the IDL library ($\tt{SVDC, LA\_SVD, IMSL\_SVD}$), moreover some of the routines that use the thread pool, such as the {\tt FFT} (actually Discrete Fourier Transform {\tt DFT}) are not performing fast as freely available solutions such as the {\tt fftw} the Fastest Fourier Transform in the West~\cite{FFTW05}. The IDL language is optimized to work with vectors and array but in some context a step backward to old fashion {\tt for} loop is necessary and these cases critically slow down the simulation speed.

As the technological development lowered the prices and pushed the performance of general purpose graphic processors (GPGPUs) these became a second way to approach to High Performance Computing (HPC). Following 
this approach the availability of mathematical/physical libraries for numerical computation has boomed. In particular we were interested into the use of a GPU to accelerate the numerical computation. \\
NVIDIA set a new standard with TESLA based on the NVIDIA Kepler\texttrademark Architecture for scientific computing, thanks also to the development of the Compute Unified Device Architecture\cite{CUDA} (CUDA). 
NVIDIA created a parallel computing platform for the GPUs they produce, which is almost perfectly compatible with the full C standard. Actually it runs code through a C++ compiler.\\
From the IDL environment it's possible to call external compiled tools stored in shared object ({\tt .so}) files on Linux operative systems or Dynamic-link library ({\tt .dll}) on Windows one, using the 
External Linking through {\tt CALL\_EXTERNAL} and, more user friendly, the dynamically loadable modules ({\tt DLM}). In particular in this way we may have access to parallel optimized C/C++ and CUDA 
libraries on IDL.

\begin{figure}
   \begin{center}
   \begin{tabular}{c}
   \includegraphics[height=8cm]{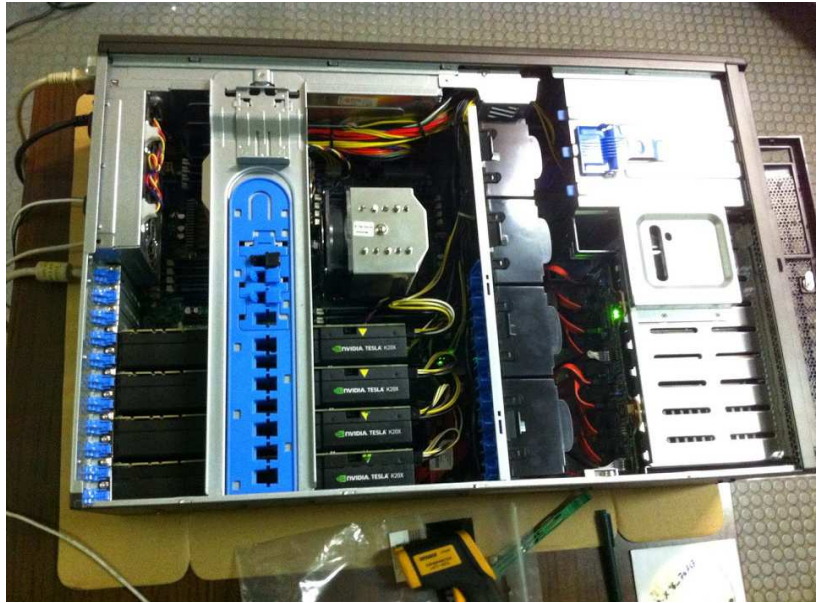}
   \end{tabular}
   \end{center}
   \caption[dyno] 
   { \label{fig:dyno} 
A picture of the workstation as seen from the side. On the bottom left the 4 Tesla GPUs}
   \end{figure}

These reasons pushed us to consider the acquisition of multi-core, multi-gpu workstation composed in this way:
\begin{itemize}
\item 2 Processors Intel Xeon E5-2630 v2, 2.6 GHz frequency (each one has 6 cores);
\item 256 GB RAM on 16 GB group;
\item 4 GPU Nvidia TESLA K20X (6 GB, 2688 CUDA cores each);
\item 4 Hard Disk, 2 TB each for data storing.
\end{itemize}
The workstation runs Fedora 19\cite{fedora} operative system, the CUDA 6.0 version and IDL 8.3, you may have a look of the GPU card in the figure~\ref{fig:dyno}.

Regarding the CUDA and the GPUs: we aim to write our CUDA library, but meanwhile we will get the skills to do it (properly) we are using the 
GPULib from the Tech-X company. It enables users to access HPC by loading a DLM into the IDL\cite{techx} session.

At the moment we are writing this proceeding we are studying the best way to make a parallel use of the GPUs. To be notice that these are mounted on two INTEL PCI~Express~2.0 cards (PCIe $\times$16), these points may limit the fast transfer of data between GPUs for parallel GPU computation.
\section{Block Diagram} 
\label{sec:scheme}
\begin{figure}
   \begin{center}
   \begin{tabular}{c}
   \includegraphics[height=9cm]{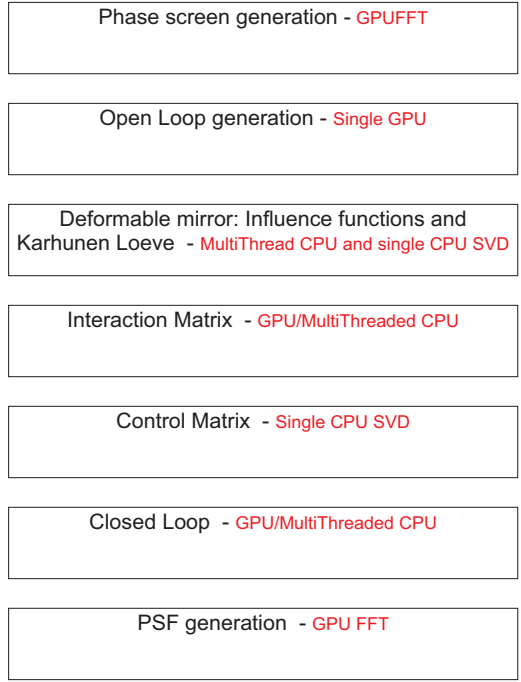}
   \end{tabular}
   \end{center}
   \caption[dyno] 
   { \label{fig:scheme} 
The different blocks composing the full E2E simulation. Each block is independent one another and the results of
each of these steps is saved in a fits file on the disk. This file is the input for the following module, if a precomputed input is needed.
}
   \end{figure}

We decomposed the full simulation into stand alone phases according to the logic of the adaptive loop itself.
Some of the ``devices'' of an E2E simulation are actually independent from the closed loop (interaction matrix) or even from the adaptive
optics instrument simulated (for example the phase screens).\\
The code we designed takes advantage of this characteristic considering the following steps as standing alone 
simulations (we call them blocks), see figure~\ref{fig:scheme} :
\begin{itemize}
\item {\bfseries The atmospheric layers generation}: we compute the phase screen re-using some of the LOST routines, partly of completely
adapted to use the GPULib, in particular the phase screens are computed as FFT from the Kolmogorov\cite{1941DoSSR..30..301K,1941DoSSR..32...16K} or Von Karman\cite{Winker:91}
models of the phase Power Spectrum,
the phase screens arrays are finally saved on a {\tt .fits}\cite{fits} file, in order to be easily re-used and easily inspected. 
\item {\bfseries The open loop wave-front measurements}: before the starts of the closed loop operation we have ready the open loop wavefront (phase) measurements history for the stars and references
we considered both for AO loop control and performance estimation over the field. The code computes the loop phase history series and it saves it on a {\tt .fits} file for the user defined angular directions and the reference stars.
\item {\bfseries Influence function and modal base}: the deformable mirrors are computed as the linear combination of a modal base. The actuators base defined by the measured or expected influence functions
is the starting base to compute a modal base for the interaction matrix and close loop phases. The tool accepts user-supplied influence functions or computes those analytically. They can be used to 
fit Zernike modes or to compute the Karhunen-Loeve\cite{karhunen} expansion which best fits the stochastic wave--fronts induced by atmospheric turbulence. Also in this case it's is to inspect the computed modes since they are stored in a (large) {\tt .fits} file.
\item {\bfseries The calibration of the interaction matrices}: this step runs completely independently from the other blocks and it only needs the proper definition of the telescope 
pupil to be consistent with the saved open loop. The modes defined in the base in input are used to register the slope measured by the WFS modules.
\item {\bfseries Control Matrix}. It is mainly the (pseudo) inverse of the array built using the DM - WFS interaction matrices.
\item {\bfseries The closed loop}: here the real action takes place. The WFS modules compute the vector signals (X- and Y- slopes) from the residual phase (the open loop wavefront subtracted by the DM).
Vector slopes multiplied by the control matrix returns the vector coefficients for the modal base linear composition, in the form of the differential coefficients to be added to the one defining the 
existing mirror shape. It's possible to add the telescope induced vibration in the Open Loop WF of the stars/references. It's possible to consider at this level the sodium layer temporal evolution. At this stage we consider a pure integrator scheme for the 
loop control.
\item {\bfseries The Point Spread Functions (PSFs)}: the generation of the PSF is computed, if needed, at the user-desired wavelength, or wavelengths, considering the color of the object whether is specified. From
the PSF the code computes SR and Encircled Energy in the preferred radius.
\end{itemize}
Finally a service block of routines analyzes the results and computes the interesting quantities from the Closed Loop and PSF {\tt .fits} files.

\section{Modules} 
\label{sec:components}
Important modules to be described are the following:
\begin{itemize}
\item {\bfseries Open Loop generation}. We started not considering Fresnel propagation: the code piles-up the phase screens properly
shifted to consider the angular direction of the star and temporal evolution. In the case of the LGS, the projections of the laser beam on the sodium layer is 
computed considering the spot shift induced by the tip-tilt integrated over the up-warding path.
In the code a laser guide star WF is the results of different, 1 to 5 typically, projections of the beam on as many sub-layers of the sodium layer: each projection (read also layer altitude)
is used to compute a WF considering that the pupil projection changes at different layers altitudes. The WFS in closed loop will take into account the split
of a single LGS into different sodium sub-layers to build the signal.
\item {\bfseries Influence Functions}. If the user does not provide a set of influence functions accounting for the actuators effect, then the code compute them using an
optimized version of the Thin Plate Spline method included in the {\tt MIN\_CURVE\_SURF} of IDL.
\item {\bfseries Karhunen-Loeve}. The Karhunen-Loeve base is ``empirically'' computed considering pupil shape, influence functions base and incoming turbulence
power spectrum. A large set of phase screens is computed and the modal fit with the actuators base is computed. The resulting covariance matrix of the coefficient vector
is averaged over the phase screens realization. This matrix is decomposed in its singular values and the right-singular vectors (by the way, the covariance matrix is symmetric) 
are used to compute the modal base once projected back on the actuators space.
\item {\bfseries LGS Shack Hartmann wavefront sensor}. We divided the sodium layer into different sub-layers. The LGS wavefront sensor module has as input the WFs computed in open loop 
subtracted by the DMs phase for each sodium sub-layer and the directions and positions of the launching telescopes. We simulate a Shack-Hartmann (SH) which computes the signal corresponding to the sodium beam. The WF is split into sub-apertures
and the spot computed by FFT for each sensor sub-aperture. For each sub-layer the proper phase shift in Tip Tilt and Defocus is considered and the spot array multiplied by the corresponding value of the sodium thickness level profile. For each of sub-layer we consider the elongation effect
in order to finally compose and image taking into account the whole layer thickness. We consider the physical extension of the projection of the laser on the sodium layer through a convolution for a gaussian of the desired properties. Different convolution model are in our developing plan.
Finally the sub-layer arrays are piled-up and the centroid computed on the final noise-added image to retrieve the slopes.
\item {\bfseries Pyramid WFS}. We follow the scheme presented in (Verinaud 2004)\cite{Verinaud200427}, see also\cite{2003SPIE.4839..524V}. This module makes a large use of the GPU computing power for the whole computation including matrix-matrix complex
array multiplications, Fast Fourier Transform and other operations on arrays.
\end{itemize}
\section{Examples} 
\label{sec:examples}
\begin{figure}
   \begin{center}
   \begin{tabular}{c}
   \includegraphics[height=5.2cm]{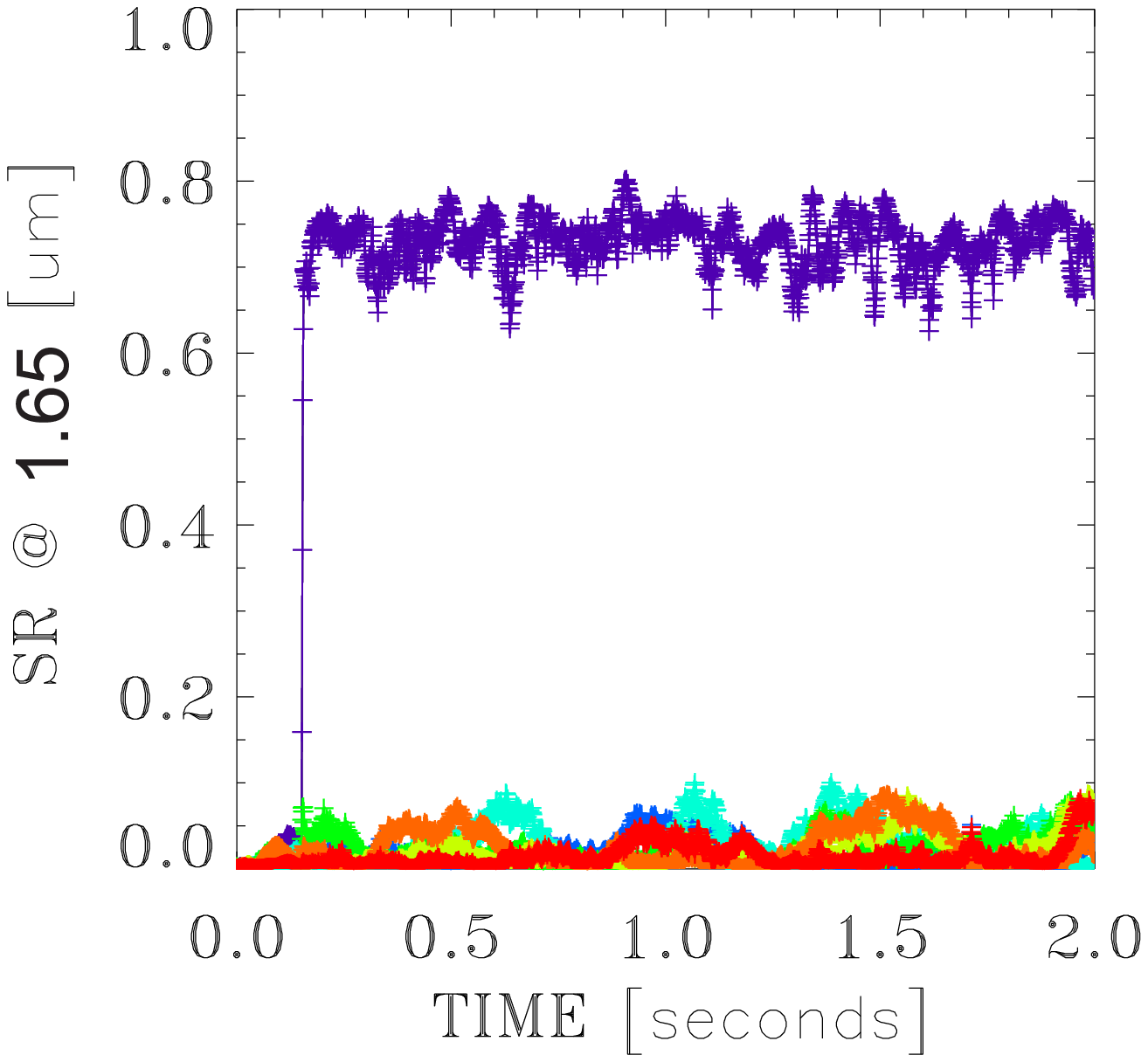}
	 \includegraphics[height=5.2cm]{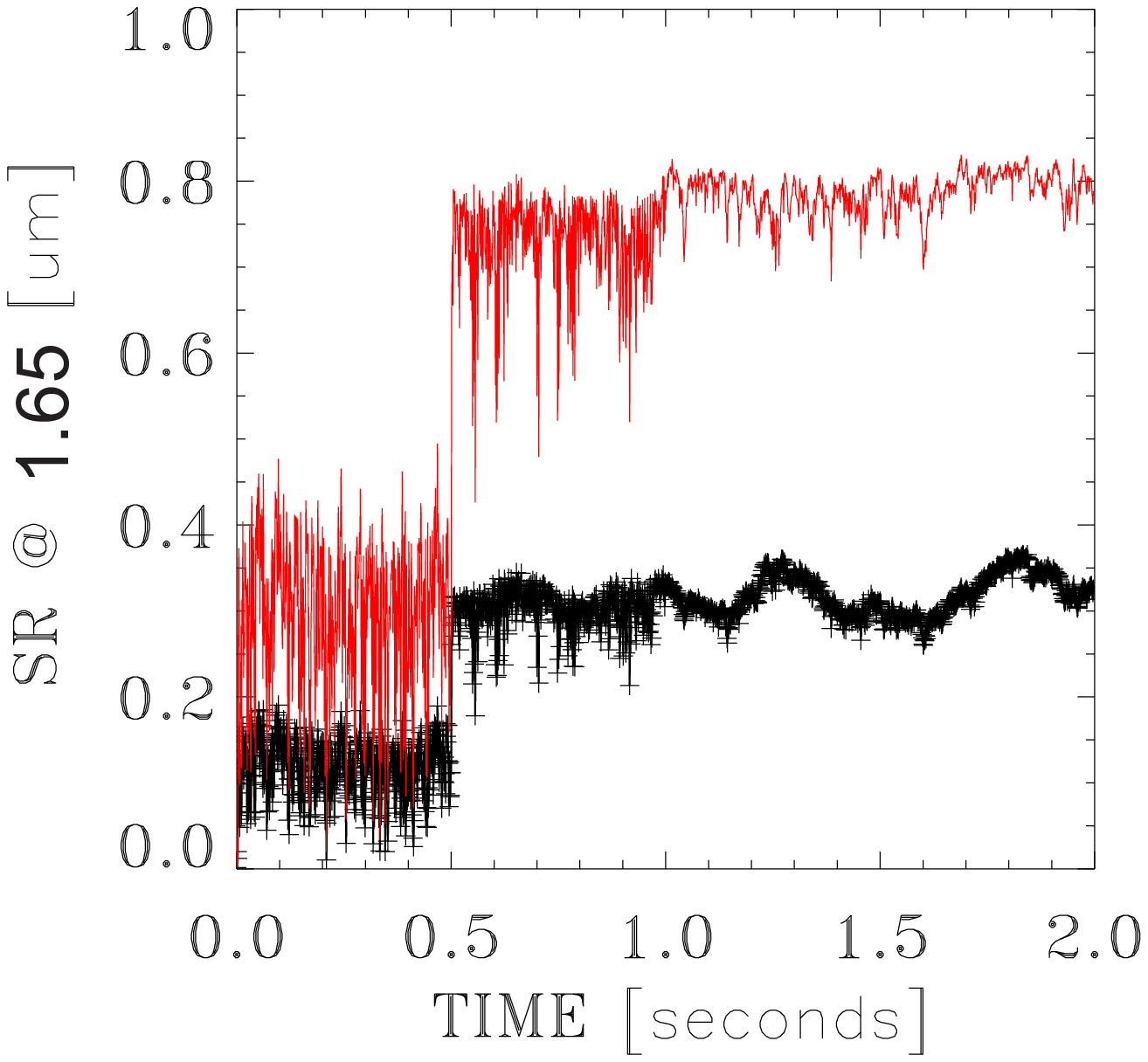}
	 \includegraphics[height=5.2cm]{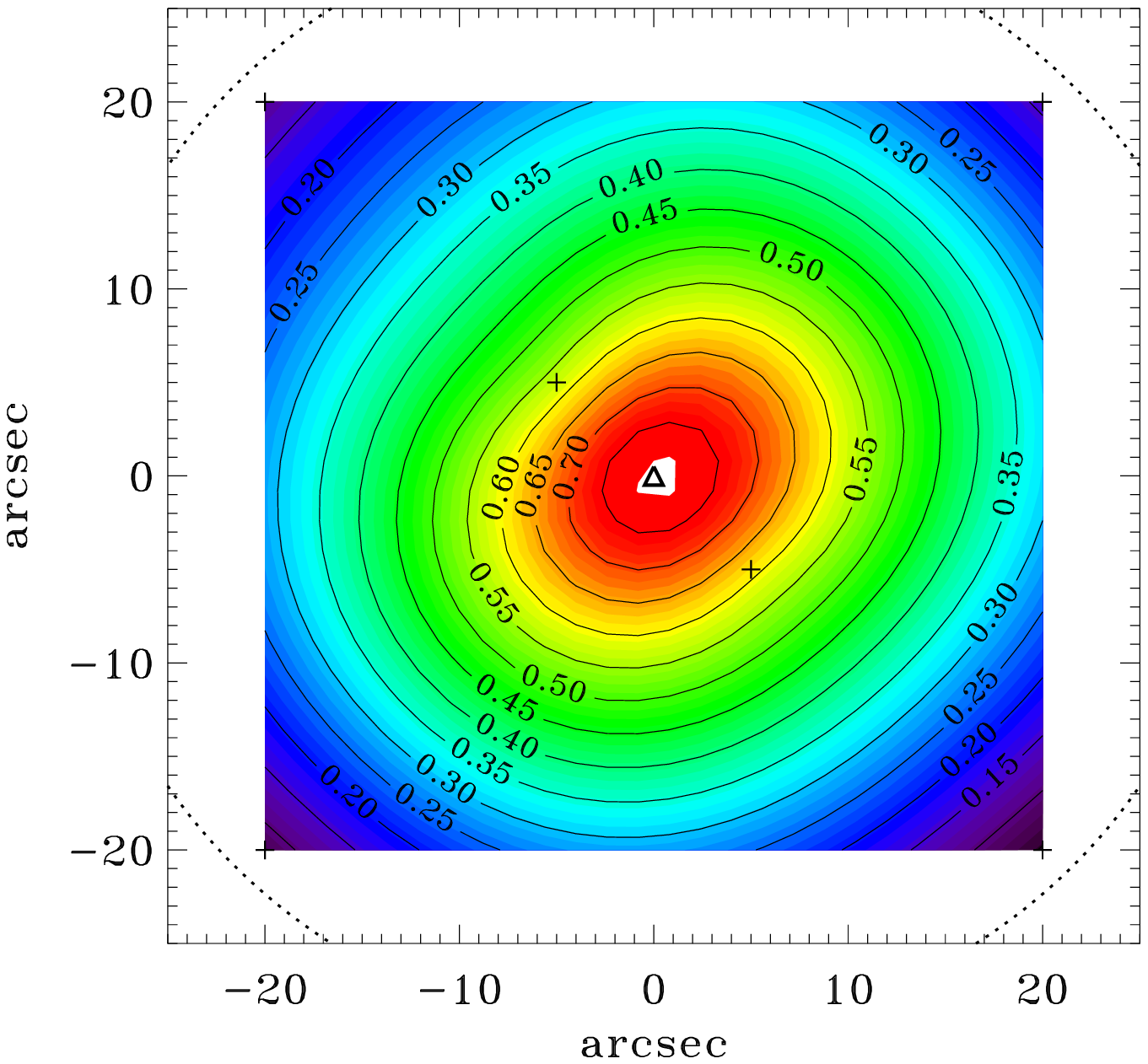}
   \end{tabular}
   \end{center}
   \caption[SR] 
   { \label{fig:sr} 
The first two plots from left to right show how the effective gain of the pyramid changes with performance (correction residuals) and how affects it (see text), 
the highest performance curve refers to the guide star, the lower performance curves to some of the field stars and to the average over the field, respectively for the left and right figure. The last picture shows the SR map variation over the field. The PSFs were computed in H band and the SR refers to this band too.
}
   \end{figure}

\begin{figure}
   \begin{center}
   \begin{tabular}{c}
   \includegraphics[height=5.2cm]{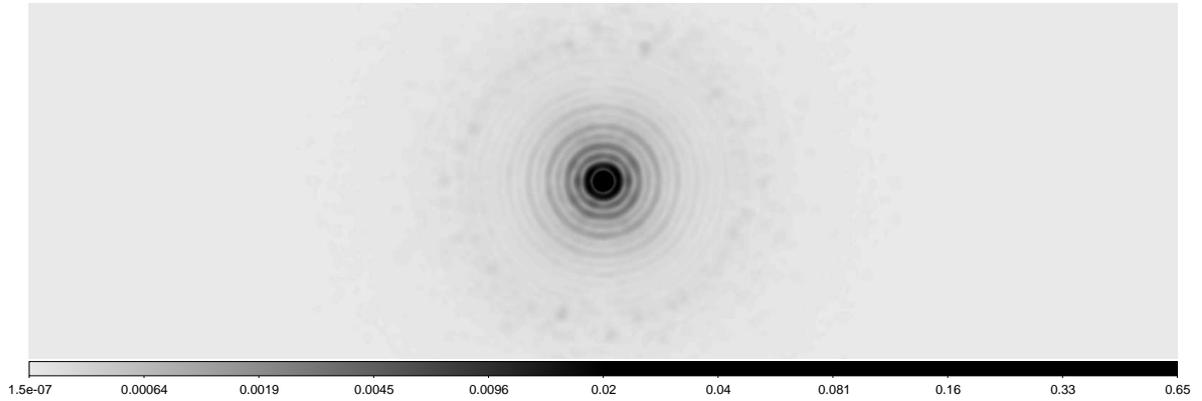}
   \end{tabular}
   \end{center}
   \caption[PSF] 
   { \label{fig:psf} 
	The picture shows the integrated PSF of the reference star of the second case above, as finally computed from the close loop residual WF series. In this case the simulated pyramid WFS and adaptive secondary mirror on the LBT telescope were retrieving a SR$\approx0.8$ in H band.
}
   \end{figure}

We tested the validity of the code whenever possible using typically the LBT telescope as test case. This allows us to compare the results obtained by the new tool to the ones obtained in 
laboratory\cite{2010ApOpt..49G.174E}, on sky\cite{FLAO} and simulated\cite{2014ExA...tmp...10A}.
As example we describe here an experiment we are running for the determination of the variation of the optical loop gain of the pyramid.
The pyramid sensitivity depends on the size of the reference star on the pyramid pin, larger this size smaller the sensitivity (with a larger linear regime range). 
On the test case we considered a 400 modes correction using a modulated ($3\lambda/D$) pyramid on a seeing of 1~arcsec in photon noise
dominated regime (5 magnitude equivalent star) and with different gain vectors applied. We simulated two scenarios:
\begin{itemize} 
\item with a fixed vector gain computed as half of the one actually calibrated after that convergence of the loop was reached with an initial gain of 1.0 for all the modes;
\item with half of the actual gain vector calibrated in open loop and waiting 0.5~sec we made a new gain estimation, again after 1.0~sec and 1.5~sec (at 0., 0.5, 1.0 and 1.5 seconds of the simulation).
\end{itemize}
In these cases the pyramid response was calibrated mode by mode: for each one we obtained a coefficient given by the ratio expected/measured value of the injected mode amplitude, see figure~\ref{fig:sr}. We calibrated in the first case after about 0.1 second once the convergence was reached with a low unitary gain for all the modes. Please notice that we get a vector of gains and not a single coefficient since we calibrated the pyramid response mode by mode. The original interaction matrix registered and used to compute the control matrix was calibrated using a modal push-pull scheme and with no aberrations (only the mode calibrated) from correction residual. During the test we have the residual phase of the correction on top of the mode to be calibrated lowering the pyramid sensitivity and the measured slopes amplitude.  Before the convergence was reached the gain vector computed were systematically too high, generating oscillations in the loop. This oscillations were present nevertheless we were applying just half of the gain computed in the calibration. For sake of clarity the calibration was performed stopping the loop and using the residual phase of the correction achieved, as it was a static aberration. 
In the second case we calibrated the modal response every 0.5seconds: only after two iterations of this calibration we succeeded to get a stable correction. The SR in the figure~\ref{fig:sr} was computed on the instantaneous PSF, which integral is presented in figure~\ref{fig:psf}.
	
	At the moment we are writing this proceeding the most time demanding module of the close loop is the LGS SH sensor which takes about 4.5sec to retrieve slopes from the incoming WFs using a parallel CPUs approach (using MAORY configuration and 3 sodium sub-layers). Considering for MAORY 3 SH sensors the time required is about 15sec. This may be enough to run a full 2seconds simulation in less than 1 day, but may be improved further. For example, considering what we already get: we completed a full calibration of the M4 mirror using one LGS in about 14hours (5000 modes, $80\times80$ subapertures) corresponding to 0.1 Hz per mode and using a push-pull modal approach.
	
\section{Conclusions and future upgrades} 
\label{sec:conclusions}
At the moment we are writing this proceeding the code is able to provide E2E MC simulation of Ground Layer Adaptive Optics (GLAO) systems using Natural and Laser guide stars on the E-ELT dimension. We have to speed up soon the Singular Value Decomposition algorithm. This has to be done on CPU 
since the matrix to be decomposed may be large as and even more than the available memory on the GPU card. \\
The translation to CUDA/GPU may reduce significantly the computation time for the LGS SH sensors to less than once per 2sec (this is our goal).
Generally speaking we want to translate the routines from the GPULib library to CUDA in order to optimize performance and minimize the calls from the IDL session to the GPUs.
The next step would be the multi-GPU approach that can boost part of the computation of a factor 4 (as the number of GPU cards available).
To be explored the translation to C/C++ of some of the more complex routines now written in IDL using the OpenMP for multithreading programming.\\
Once the code will be debugged and it achieve a reasonable fast speed we will move to more complex cases completing the MCAO LGS system of MAORY.
	
\subsection{Acknowledgments} 
This work has been partly supported by the Italian Ministero dell’Istruzione, dell’Universit\`a e della 
Ricerca (Progetto Premiale E-ELT 2012 – ref. Monica Tosi).  
 %
%
%
\bibliography{biblio}   
\bibliographystyle{spiebib}   

\end{document}